\documentclass[a4paper,11pt]{article}
\pdfoutput=1 

\usepackage{jcappub} 
\usepackage{braket} 

\usepackage[T1]{fontenc} 
\usepackage{subcaption}

\title{\boldmath Spectator fields in the post-inflationary Starobinsky model after ACT}

\author[\dagger]{Norma Sidik Risdianto}
\author[\star]{Romy Hanang Setya Budhi}
\author[\star]{Muhammad Abdan Syakura}

\affiliation[\dagger]{Dept. of Physics Education, Universitas Islam Negeri Sunan Kalijaga,\\Jl. Marsda Adisucipto 55281, Yogyakarta, Indonesia}
\affiliation[\star]{Physics Department, Universitas Gadjah Mada, Yogyakarta 55281, Indonesia}

\emailAdd{norma.risdianto@uin-suka.ac.id}
\emailAdd{romyhanang@ugm.ac.id}
\emailAdd{muhammadabdansyakura1997@mail.ugm.ac.id}

\abstract{We investigate the post-inflationary dynamics of the conventional Starobinsky model in light of the recent ACT results, focusing on its preheating mechanism. We adopt a matter-dominated era with $w=0$, as naturally predicted by the quadratic scalaron potential near its minimum. Although this scenario lies at the edge of the ACT-LB constraints, it is excluded by P-ACT-LB, which instead favors $w\simeq 0.5$. Nevertheless, we retain $w=0$ to study preheating within the conventional Starobinsky model. We consider a two-stage preheating scenario: the first stage is driven by scalaron oscillations, while the second is dominated by the spectator field. In the first stage, tachyonic preheating provides an efficient energy transfer from the inflaton. We also investigate the spectator field as a dark matter candidate and constrain its coupling to obtain a viable scenario. Furthermore, we examine primordial black hole formation and find that it occurs predominantly during the early stages of preheating, consistent with previous studies. In the second stage, the quartic spectator-field potential drives the universe into a radiation-dominated era and determines the reheating temperature. This process involves Higgs particle production at the first zero crossing of the spectator field, followed by perturbative decay into $b\bar{b}$ pairs.}

\begin{document}
\maketitle
\flushbottom
\newpage

\section{Introduction}

The Starobinsky inflationary model \cite{starobinsky1980new,de2010f,sotiriou2010f} and its descendants \cite{kehagias2014remarks} were considered among the best-fit inflationary models based on the PLANCK results \cite{akrami2020planck}. These results predict a reheating temperature much lower than that of most other inflationary models \cite{terada2015reheating,garcia2023reheating,rodrigues2022higher}, typically around $T_\text{reh}\sim 10^8-10^9$ GeV. However, Ref. \cite{dorsch2024gravitational} showed that pure gravitational reheating through non-perturbative gravitational particle production during Starobinsky inflation preheating could lead to a maximum reheating temperature of $T_\text{reh}\sim 10^{12}$ GeV (see also Ref. \cite{van2017reheating}).
In light of the new ACT results, particularly P-ACT-LB, significant tension arises with the conventional Starobinsky inflationary model, as these data strongly disfavor it. However, ACT-LB still leaves a small allowed region in which the conventional Starobinsky model remains viable at the edge of the $2\sigma$ region, assuming an equation of state $w=0$ for $55$--$56$ e-folds. In this region, the reheating temperature is constrained to be significantly higher than previously estimated, at $T_\text{reh}\sim 10^{13}$--$10^{15}$ GeV \cite{risdianto2026preheating,mohammadi2026starobinsky}.

The reheating temperature should be directly related to the preheating stage \cite{dolgov1989particle,traschen1990particle,shtanov1995universe,kofman1994reheating}. 
During this stage, the coherent oscillation of the inflaton field leads to a rapid and non-perturbative transfer of energy into other particle species through mechanisms such as parametric resonance \cite{kofman1997towards,greene1997structure}. The efficiency of this energy transfer determines the amount of the universe's thermalization. Consequently, any modification of the preheating dynamics is expected to alter the predicted reheating temperature. However, the preheating mechanism has no direct observational signature and is therefore highly model-dependent\footnote {Likewise, the reheating temperature lacks observational constraints}. In that case, relating the preheating with other cosmological signature such as axion and leptogenesis \cite{hashimoto2021inflation}, primordial black hole (PBH) \cite{risdianto2025second,risdianto2026preheating}, dark matter (DM) \cite{risdianto2025inflation}, and the gravitational waves \cite{dufaux2007theory,garcia2007stochastic,dufaux2010gravitational,bethke2013anisotropies} could enhance its visibility.

In this paper, we investigate the preheating mechanism of the Starobinsky model. Since the scalaron potential becomes approximately quadratic near the end of inflation, the universe naturally enters a matter-dominated era with $w=0$. We therefore adopt this post-inflationary scenario, although it places the model at the edge of the ACT-LB constraints and totally excluded by P-ACT-LB results.
We then examine the reheating mechanism associated with the preheating stage and investigate how the dynamics of preheating influence the thermalization process. Particular attention is given to the connection between particle production during preheating and the resulting reheating temperature. In addition, we introduced the two-stages of preheating to solve the high-temperature prediction suggested by the ACT results. We will discuss the importance of the spectator field in both stages of preheating. 

Finally, this paper is organized as follows: in section \ref{model}, we discuss the Starobinsky model. In section \ref{postinflation}, we give the brief explanation of the post-inflation in Starobinsky model. In section \ref{firstpreheat}, we discuss the first preheating, in which the role of the spectator field, tachyonic preheating, the dark matter (DM) abundance, the primordial black hole (PBH), and the end of the first preheating are also discussed. Later, in section \ref{secondpre}, we discuss the second preheating, where the spectator field is responsible on the thermalization of the reheating temperature.

\section{The Model}\label{model}
The action in the Starobinsky model can be written as
\begin{equation}
    S_J=\int d^4x\sqrt{-g_J}\left(\frac{1}{2}M_p^2f(R_J) \right), \hspace{1cm} f(R_J)= R_J+\frac{R^2_J}{6M^2},
\end{equation}
where $R$ and subscript $J$ represent the Ricci scalar and the Jordan frame, respectively.$M$ is the effective mass of the scalaron at the end of inflation. In this paper, we set\footnote{This value is determined from the normalization of the scalar power spectrum using the P-ACT-LB best-fit values. The previous PLANCK results give $M\simeq 1.3 \times 10^{-5}M_p$, which is slightly larger than our value. However, this small difference has a negligible effect on all the calculations presented in this paper.} $M=9.07 \times 10^{-6}M_p$.
To obtain the action in the Einstein frame, we used the Weyl transformation as
\begin{equation}
    g^{\mu\nu}_E\equiv\Omega^{-2}  g^{\mu\nu}_J, \hspace{1cm}\Omega^{2}\equiv\frac{df(R_J)}{dR_J}=1+\frac{R_J}{3M^2}\equiv\exp\left({\sqrt{\frac{2}{3}}\frac{\phi}{M_p}}\right).
\end{equation}
We obtain the action in the Einstein frame as follows
\begin{equation}\label{se}
    S_E=\int d^4x \sqrt{-g_E}\left(\frac{1}{2}M_p^2 R_E -\frac{1}{2}g^{\mu\nu}_E\partial_\mu \phi\partial_\nu \phi-V(\phi) \right),
\end{equation}
where ${R}_E=\Omega^{-2}R_J-6\Omega^{-3}\square\Omega$ and

\begin{equation}\label{potentialphi}
    V(\phi)=\frac{3}{4}M_p^2M^2\left(1-e^{-\sqrt{\frac{2}{3}}\frac{\phi}{M_p}}\right)^2.
\end{equation}
For the rest of this paper, we will omit the subscript $E$ for simplicity and adopt the $(-,+,+,+)$ signature.

\section{The post-inflationary mechanism}\label{postinflation}

After the end of inflation, when the amplitude of the scalaron $\phi$ is around $\sim M_p$, the scalaron oscillates around the minimum of its potential. At this time, the potential in Eq. \eqref{potentialphi} can be approximated as
\begin{equation}\label{potentialphi2}
V(\phi)\simeq \frac{1}{2}M^2\phi^2.
\end{equation}
At this stage, the Universe is filled with heavy particles produced through resonant particle production during preheating, as discussed in the following subsection. The equation of state is approximately $w\approx 0$, which follows from the quadratic nature of the potential and favors a heavy-matter-dominated regime just after inflation. However, the conventional Starobinsky inflationary model is excluded by the P-ACT-LB results when assuming $w=0$. In Refs. \cite{drees2025refined,mohammadi2026starobinsky,zharov2025act}, the new P-ACT-LB results are shown to favor Starobinsky inflation with $w\sim 0.5$. This implies that the post-inflationary regime favored by this equation of state is a stiff-like era, in which the dominant energy component loses energy faster than radiation. This scenario could be promising for both satisfying the P-ACT-LB constraints and achieving efficient preheating. However, during this period, the potential is dominantly proportional to $\phi^2$. For $w=0.5$, the potential should instead take the form $\propto\phi^6$, which requires an extension of the conventional Starobinsky model and is beyond the scope of this paper. The main purpose of this paper is to accommodate the conventional Starobinsky model, which favors a $\propto\phi^2$ potential after inflation. Fortunately, $w\simeq 0$, corresponding to the quadratic regime described by the potential in Eq. \eqref{potentialphi2}, remains consistent with the ACT-LB results for $N\simeq 55$--$56$ e-folds \cite{risdianto2026preheating}.\footnote{Larger values of $N$ are constrained by the reheating temperature, with $T_\text{reh}\gtrsim 10^{15}$ GeV.}

\begin{figure}[h!]
    \centering
    \includegraphics[width=0.6\linewidth]{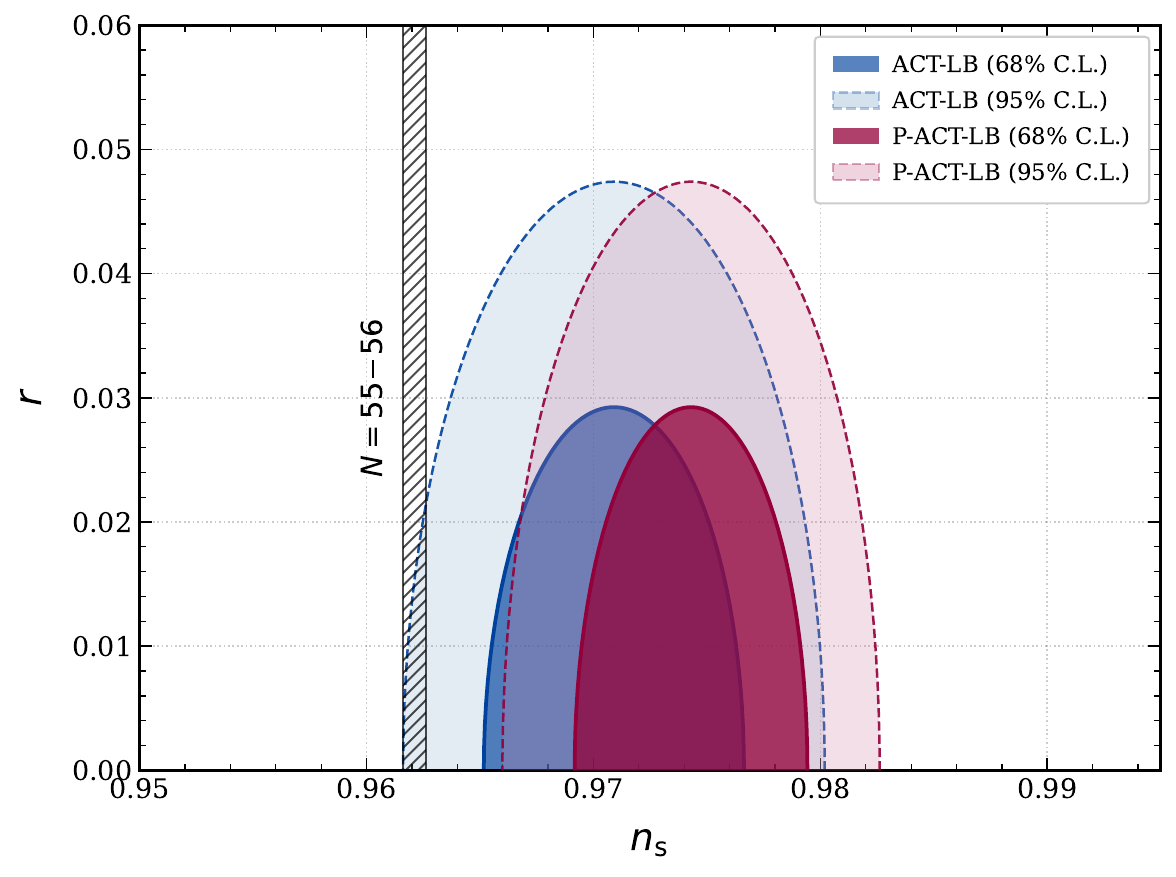}
    \caption{The contour of P-ACT-LB (red) and the ACT-LB (Blue). Our preferred regime for this paper lies in the shaded area at the edge of the ACT-LB contour.}
    \label{allowedregime}
\end{figure}

In this paper, we consider only this parameter region as the most plausible scenario in which the conventional preheating stage occurs in the Starobinsky model. Furthermore, since our study also discusses PBH production during preheating, adopting $w=0$ is particularly important, as a matter-dominated post-inflationary phase favors PBH formation during this period.
Finally, Fig. \ref{allowedregime} shows our preferred parameter region.

In this paper, the regime in which the inflaton oscillates immediately after the end of inflation is referred to as the \textit{first preheating}. This stage is followed by the \textit{second preheating}, during which the spectator field takes over the oscillation.

\section{The \textit{first} preheating}\label{firstpreheat}

\subsection{The role of the spectator field in the first preheating}
In this paper, we partially adopt the preheating scenario from Ref. \cite{risdianto2025second}, in which the preheating mechanism is divided into two stages. The details of this mechanism will be discussed shortly. As a preliminary step, we introduce the field $\chi$, whose action can be written as
\begin{equation}\label{sx}
    S_\chi=\int d^4 x \sqrt{-g_J}\left(-\frac{1}{2} \partial_\mu \chi \partial^\mu \chi-\frac{1}{2}\xi R_J \chi^2 -\frac{1}{4}\lambda \chi^4 \right).
\end{equation}
At the end of inflation, the potential in Eq. \eqref{potentialphi} becomes approximately quadratic (see Eq. \eqref{potentialphi2}), and the analytical solution for $\phi$ can be estimated as \cite{kofman1997towards}.

\begin{equation}\label{phi}
    \phi(t)=\Tilde{\phi}(t) \sin (M t), \hspace{1cm} \Tilde{\phi}(t)=2\sqrt{\frac{2}{3}} \frac{M_p}{Mt}.
\end{equation}

Straightforwardly, the full action in the Einstein frame of the $R^2$ term with the $\chi$-field can be written as
\begin{equation}\label{sechi}
\begin{split}
    S=&\int d^4x \sqrt{-g}\Bigg[ \frac{1}{2}M_p^2 R -\frac{1}{2}g^{\mu\nu}\partial_\mu \phi \partial_\nu \phi-\frac{3}{4}M_p^2M^2\left(1-e^{-\sqrt{\frac{2}{3}}\frac{\phi}{M_p}} \right)^2 \\
    & -e^{-3\sqrt{\frac{2}{3}}\frac{\phi}{M_p}}\frac{1}{2}g^{\mu\nu}\partial_\mu \chi \partial_\nu \chi- e^{-2\sqrt{\frac{2}{3}}\frac{\phi}{M_p}}\frac{\lambda}{4}\chi^4-  \frac{3}{2}\xi M^2  \left(e^{-\sqrt{\frac{2}{3}}\frac{\phi}{M_p}}-e^{-2\sqrt{\frac{2}{3}}\frac{\phi}{M_p}} \right)  \chi^2 \Bigg].
\end{split}
\end{equation}
Then, the equation of motion of  $\chi$ can be obtained as

\begin{equation}\label{ddotchi}  
\begin{split}
        &\Ddot{\chi}_k+3H\Dot{\chi}_k-3\sqrt{\frac{2}{3}}\frac{\Dot{\phi}\Dot{\chi}_k}{M_p}\\
        &\hspace{1cm}+\left[\frac{k^2_\chi}{a^2}   +3e^{\sqrt{\frac{2}{3}}\frac{\phi}{M_p}}\lambda \braket{\chi^2}+ 3\xi M^2  e^{2\sqrt{\frac{2}{3}}\frac{\phi}{M_p}}\left(1-e^{-\sqrt{\frac{2}{3}}\frac{\phi}{M_p}}\right) \right] \chi_k=0, 
\end{split}
\end{equation}
where we used the Heisenberg representation as
\begin{equation}\label{heisenberg}
    \chi(\textbf{x},t)=\int \frac{d^3k}{(2\pi)^{\frac{3}{2}}}\left(\hat{a}_k \chi_{k}(t)e^{-i\textbf{k}_\chi\cdot \textbf{x}} +\hat{a}^\dagger_k \chi^*_{k}(t)e^{i\textbf{k}_\chi\cdot \textbf{x}}\right).
\end{equation}
In addition, to fix the anomalous kinetic term, we will shift $\chi_k\rightarrow \chi_k\exp\left({\frac{1}{\sqrt{6}}\frac{\phi}{M_p}}\right)$ and obtain

\begin{equation}\label{ddotchik}
\Ddot{\chi}_k+3H\Dot{\chi}_k +\left[\frac{k^2_\chi}{a^2}   +3e^{2\sqrt{\frac{2}{3}}\frac{\phi}{M_p}}\lambda \braket{\chi^2}+ 3\xi M^2  \left(1-e^{-\sqrt{\frac{2}{3}}\frac{\phi}{M_p}} \right) -\frac{\Dot{\phi}^2}{6M_p^2}+\frac{\Ddot{\phi}+3H\Dot{\phi}}{\sqrt{6}M_p}\right]\chi_k =0.
\end{equation}

\begin{figure}[htbp]
    \centering
    \begin{subfigure}[t]{0.32\textwidth}
        \centering
        \includegraphics[width=\linewidth]{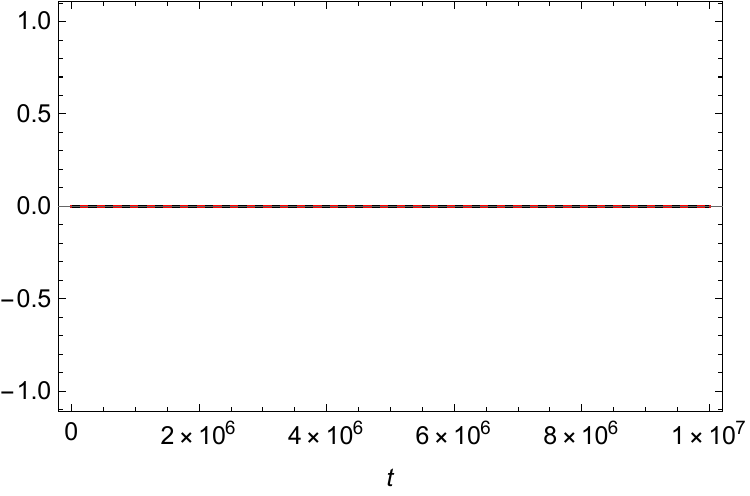}
        \caption{$\chi_k(0)=0$, $\xi=1$}
    \end{subfigure}
    \hfill
    \begin{subfigure}[t]{0.32\textwidth}
        \centering
        \includegraphics[width=\linewidth]{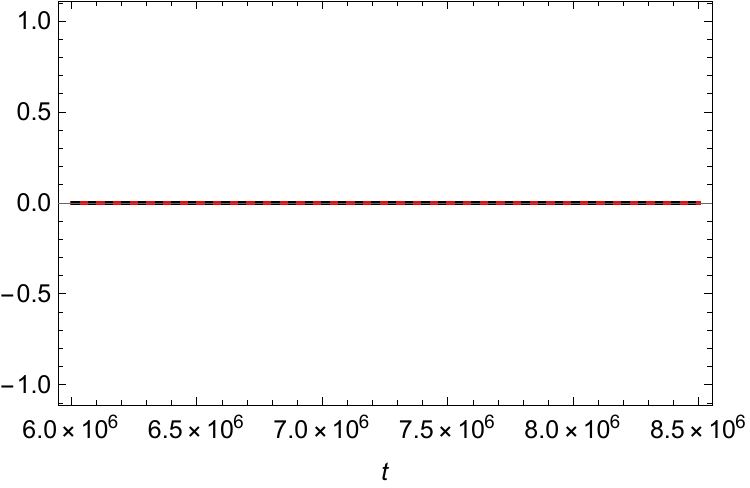}
        \caption{$\chi_k(0)=0$, $\xi=5$}
    \end{subfigure}
    \hfill
    \begin{subfigure}[t]{0.32\textwidth}
        \centering
        \includegraphics[width=\linewidth]{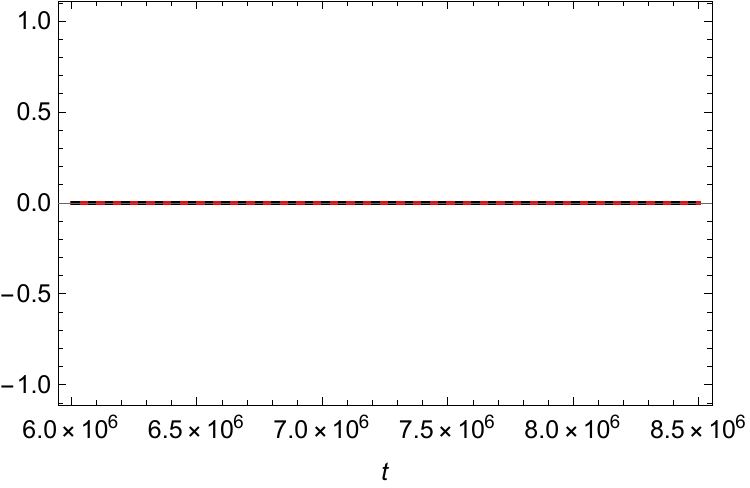}
        \caption{$\chi_k(0)=0$, $\xi=10$}
    \end{subfigure}
    
    \vspace{0.5cm}
    
    \begin{subfigure}[t]{0.32\textwidth}
        \centering
        \includegraphics[width=\linewidth]{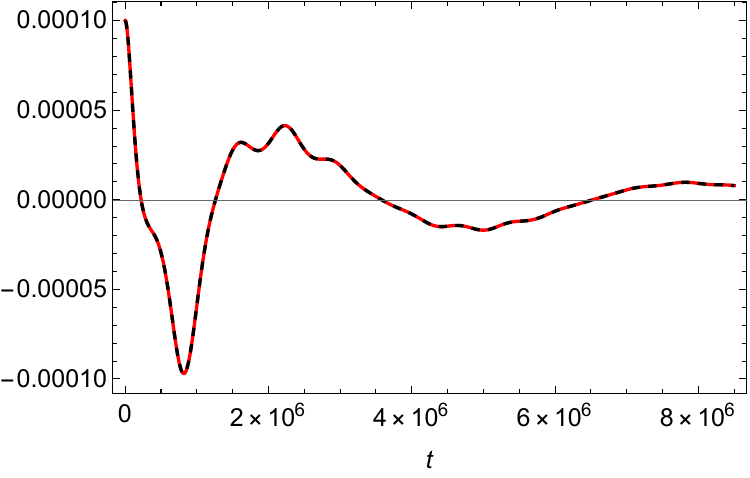}
        \caption{$\chi_k(0)=10^{-4}M_p$, $\xi=1$}
    \end{subfigure}
    \hfill
    \begin{subfigure}[t]{0.32\textwidth}
        \centering
        \includegraphics[width=\linewidth]{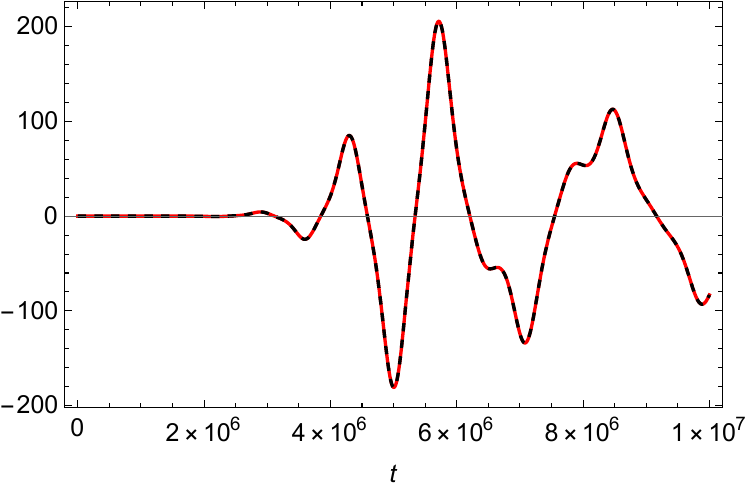}
        \caption{$\chi_k(0)=10^{-4}M_p$, $\xi=5$}
    \end{subfigure}
    \hfill
    \begin{subfigure}[t]{0.32\textwidth}
        \centering
        \includegraphics[width=\linewidth]{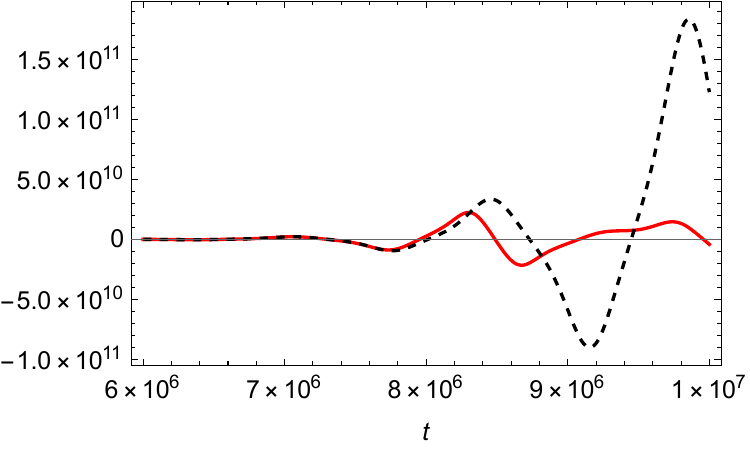}
        \caption{$\chi_k(0)=10^{-4}M_p$, $\xi=10$}
    \end{subfigure}

    \vspace{0.5cm}

    \begin{subfigure}[t]{0.32\textwidth}
        \centering
        \includegraphics[width=\linewidth]{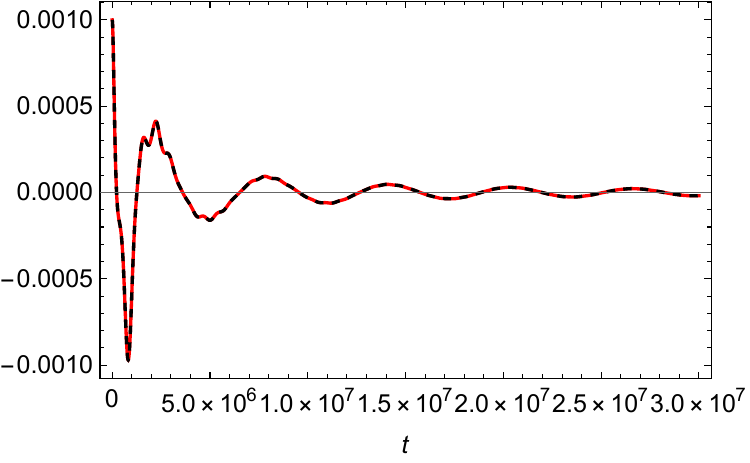}
        \caption{$\chi_k(0)=10^{-3}M_p$, $\xi=1$}
    \end{subfigure}
    \hfill
    \begin{subfigure}[t]{0.32\textwidth}
        \centering
        \includegraphics[width=\linewidth]{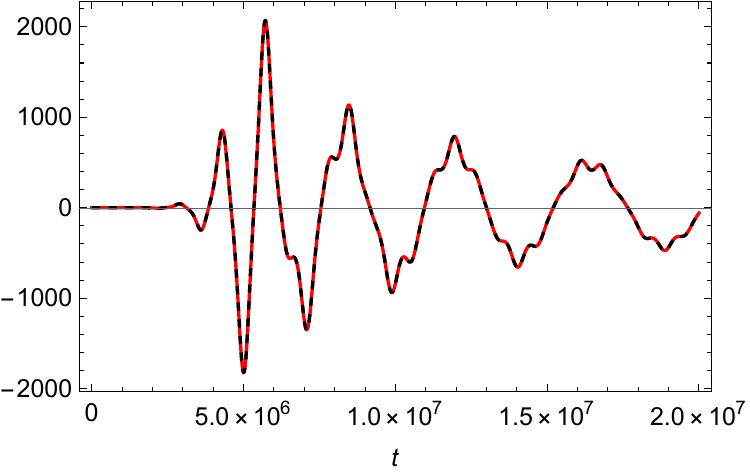}
        \caption{$\chi_k(0)=10^{-3}M_p$, $\xi=5$}
    \end{subfigure}
    \hfill
    \begin{subfigure}[t]{0.32\textwidth}
        \centering
        \includegraphics[width=\linewidth]{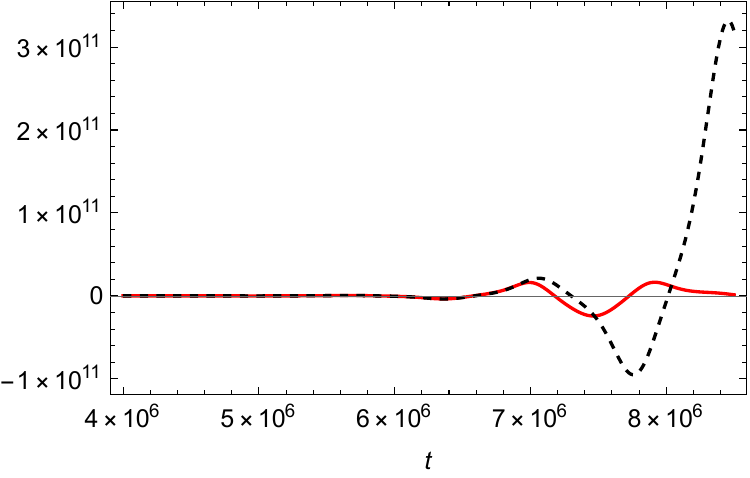}
        \caption{$\chi_k(0)=10^{-3}M_p$, $\xi=10$}
    \end{subfigure}

    \caption{The growth of $\chi_k$ with (solid-red) and without (dashed) backreaction. For all panels we used $k_i=10^{-6}M_p$, $\Delta k=10^{-12}M_p$, $\lambda=10^{-6}$.}
    \label{figure1}
\end{figure}

In Fig. \ref{figure1}, we show the growth of $\chi_k$ by solving its equation of motion using Eqs. \eqref{ddotchik} and \eqref{phi}. In this numerical calculation, we evaluate
\begin{equation}
    \braket{\chi^2}=\sum \frac{k_i^2\Delta k}{2\pi^2}|\chi_k|^2,
\end{equation}
where
\begin{equation}
    \Delta k=\frac{k_\text{max}-k_\text{min}}{N_k},
\end{equation}
and $k_i$ represents the momentum of the $i$-th mode. Here, $k_\text{max}$ and $k_\text{min}$ represent the maximum and minimum values of $k$, respectively, and $N_k$ is the total number of momentum modes. Although $\braket{\chi^2}$ is enhanced by the factor $\exp\left(2\sqrt{2/3}\phi/M_p\right)$, the fluctuations during inflation are heavily suppressed, making this term naturally small at that time.

The numerical results in Fig. \ref{figure1} show that, regardless of the value of $\xi$, no growth occurs if the initial condition for $\chi_k$ is set to zero. The energy transfer to the daughter fields then relies solely on the perturbative decay of the scalaron, which is generally considered disfavored \cite{felder1999inflation}, as it does not provide an efficient mechanism for draining the inflaton energy. In the same figure, we vary the initial conditions and the non-minimal coupling $\xi$ and find that both can trigger particle production. Nevertheless, nonzero initial conditions for $\chi_k$ are required to obtain its growth. These initial conditions may arise from several sources, such as inflaton decay, scattering with the scalaron\footnote{Both decay and scattering are expected to be heavily suppressed during inflation, but they can still generate nonzero initial values of $\chi$ at the onset of preheating.}, quantum fluctuations, or the presence of spectator fields during inflation. However, an initial condition of the order of $10^{-3}M_p$ at the end of inflation is most likely associated with a spectator field \cite{risdianto2026preheating}. In addition, Fig. \ref{figure1} shows the growth of $\chi_k$ with (solid red curves) and without (dashed curves) backreaction. The numerical results show that the backreaction effect strongly depends on the non-minimal coupling $\xi$, although this dependence is not explicitly apparent in Eq. \eqref{ddotchik}.

In the following discussion, we estimate the energy density at each zero crossing. Since our main purpose is to determine whether preheating is efficient, a rough estimate will suffice. To calculate the energy density at each crossing, we first set $a=1$ and assume $\tilde{\phi}\ll M_p$ in Eq. \eqref{ddotchik}, obtaining
\begin{equation}\label{ddotchikk}
    \Ddot{\chi}_k+\Bigg[ k^2_\chi+3\lambda\braket{\chi^2}  + 3\sqrt{\frac{2}{3}}\frac{1}{M_p}\left(\lambda\braket{\chi^2}
    +\left(\xi-\frac{1}{6}\right) M^2  \right)\tilde{\phi}\sin(Mt)  \Bigg] \chi_k\simeq 0.  
\end{equation}
Near the zero crossing, where $\sin(Mt)\approx Mt$, we define
\begin{equation}
\begin{split}
    & K^2=\left(k_\chi^2+3\lambda\braket{\chi^2}\right) \left[3\sqrt{\frac{2}{3}}\frac{1}{M_p}\left(\lambda\braket{\chi^2}
    +\left(\xi-\frac{1}{6}\right) M^2  \right)\tilde{\phi} M\right]^{-2/3}, \\
    &\eta=\left[3\sqrt{\frac{2}{3}}\frac{1}{M_p}\left(\lambda\braket{\chi^2}
    +\left(\xi-\frac{1}{6}\right) M^2  \right)\tilde{\phi} M \right]^{1/3}t.
\end{split}
\end{equation}
The equation of motion then becomes
\begin{equation}
    \frac{{d^2\chi}_k}{d\eta^2}+\left( K^2+\eta \right) \chi_k= 0.  
\end{equation}
The particle production can be estimated as
\begin{equation}\label{deltarho}
    \delta \rho_{\chi} = \int^\infty_0 \frac{d^3k_\chi}{(2\pi)^3}e^{-\pi K^2} |\Bar{m}|,
\end{equation}
where
\begin{equation}
\Bar{m}= \sqrt{  3\sqrt{\frac{2}{3}}\frac{1}{M_p}\left(\lambda\braket{\chi^2}
    +\left(\xi-\frac{1}{6}\right) M^2  \right)\tilde{\phi}  }.
\end{equation}
Finally, we obtain the energy density of $\chi$ for each inflaton zero crossing as
\begin{equation}\label{rhochi}
    \delta \rho_\chi\approx 0.27  M^4 \left(\frac{\xi\tilde{\phi}}{M_p} \right)^{3/2}
\end{equation}
The analytical calculation shows that the energy density produced during the zero crossing is much smaller than the inflaton energy density at the end of inflation, $\rho_\phi \simeq \frac{1}{2} M^2 \tilde{\phi}_\text{end} \sim 10^{-11} M_p^4$. Specifically, the energy density produced during the zero crossing is approximately ten orders of magnitude smaller than the inflaton energy density.\footnote{We assume that the calculation is performed at the beginning of preheating, when the backreaction effect $\braket{\chi^2}$ is still small.} This indicates that particle production through the zero-crossing mechanism in the Starobinsky inflation model is inefficient. Furthermore, the rough estimate in Eq. \eqref{rhochi} is expected to be at least two orders of magnitude larger than the actual numerical result because we take $a=1$ \cite{risdianto2026preheating}. Nevertheless, Eq. \eqref{rhochi} provides a rough estimate indicating that preheating through this mechanism is extremely inefficient. In the following subsection, \ref{tachyon}, we show that the efficient preheating observed in the numerical results in the last panel of Fig. \ref{figure1} is instead driven by tachyonic preheating.

\subsection{The tachyonic preheating}\label{tachyon}

\begin{figure}
    \centering
    \includegraphics[width=0.5\linewidth]{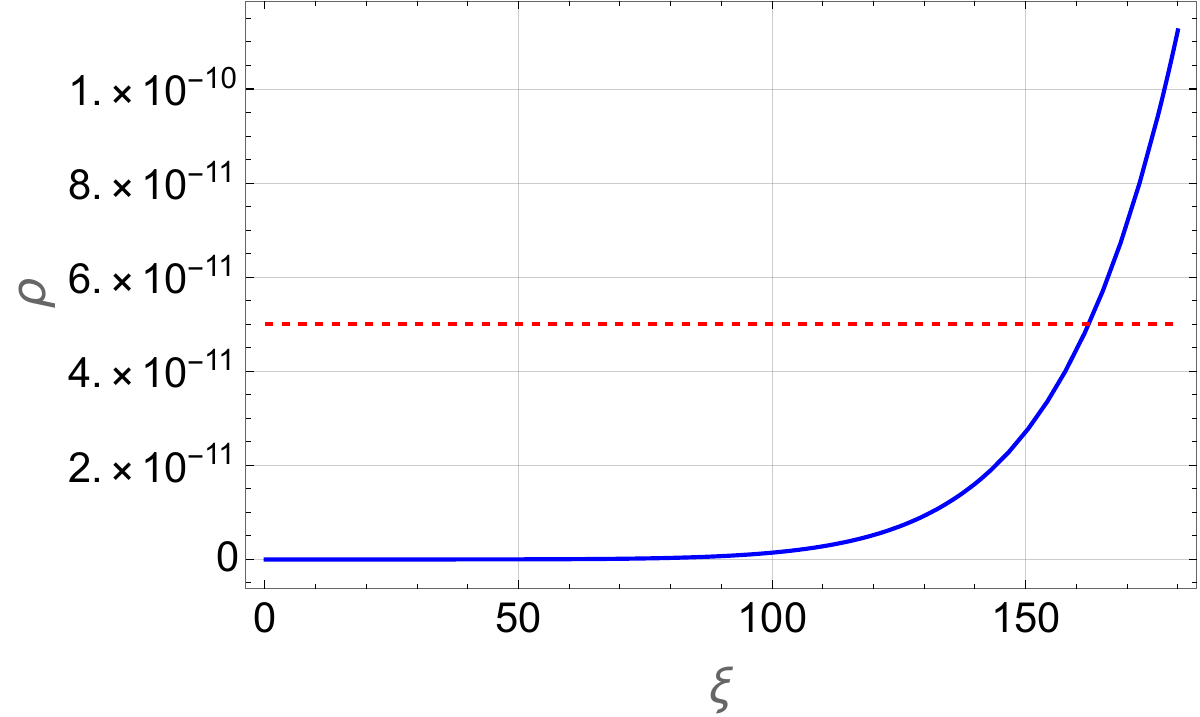}
    \caption{The inflaton energy density at the end of inflation (red dashed line) and the energy density produced by tachyonic preheating (solid blue line) as functions of the non-minimal coupling parameter $\xi$. The energy densities are expressed in units of $M_p^4$.}
    \label{plottachyon}
\end{figure}

In this section, we discuss the analytical description of tachyonic preheating, which drives the efficient energy transfer from the inflaton field. The particle production due to tachyonic preheating can be estimated as
\begin{equation}\label{rhotachy}
    \delta \rho_\text{tach}\simeq\int^{\bar{m}}_0 \frac{dk_\chi}{2\pi^2} k^2_\chi e^{\pi \cdot \frac{k_\chi^2}{(\bar{m}^2M)^{2/3}}}\bar{m} =\frac{(\bar{m}^2M)^{2/3}\bar{m}^2}{4\pi^3} e^{\pi\cdot \frac{\bar{m}^2}{(\bar{m}^2M)^{2/3}}}.
\end{equation}
In Fig. \ref{plottachyon}, we see that tachyonic preheating could potentially drain the entire energy density of the inflaton field. However, this is only a rough estimate, as we neglect the expansion of the universe and the momentum-mode distribution. Nevertheless, Fig. \ref{plottachyon} and the energy-density estimate in Eq. \eqref{rhotachy} serve to illustrate that efficient energy transfer from the inflaton field in Starobinsky inflation can be achieved through tachyonic preheating. An exact calculation would require solving the coupled nonlinear equations of motion and possibly performing lattice simulations, which are beyond the scope of this paper. Finally, tachyonic preheating can be efficient for a non-minimal coupling of $\xi \gtrsim \mathcal{O}(2)$.

\subsection{The dark matter abundance}\label{dmabundance}
In Ref. \cite{risdianto2025inflation}, dark matter (DM) is introduced as the dominant particle responsible for draining the inflaton energy in an extension of the Starobinsky model. This scenario is interesting because it naturally explains for the present-day DM abundance. In this work, we also consider $\chi$ as a DM candidate. In the following, we investigate the resulting DM abundance.

The DM in this model has an interaction term with the Higgs field via 
\begin{equation}\label{dm2hh}
    \frac{1}{4}g\chi^2h^2,
\end{equation}
where $h$ is the Higgs field in the unitary gauge. Straightforwardly, the DM abundance can be derived from the Boltzmann equation to obtain \cite{risdianto2025inflation}
\begin{equation}
    \Omega \mathtt{h}^2 \approx \frac{1.5 \times 10^{-10}}{ \sigma_{\chi\chi\rightarrow hh\nu}\text{GeV}^{-2}},
\end{equation}
where we used $\Omega \mathtt{h}^2=0.118$ \cite{louis2025atacama}. Finally, it gives a strong constraint on the DM coupling $g$ and mass $m_\chi$ as
\begin{equation}\label{mchig}
    \frac{m_\chi}{g}\approx 2.8  \hspace{1mm}\text{TeV}.
\end{equation}
As $m_\chi$ should be higher than the Higgs field\footnote{DM mass can be somewhat smaller than the Higgs field. However, in our model, it is suppressed kinematically, so $m_\chi\geq m_h$ is required.}  which favors the heavy mass of $\chi$ with a coupling range of $0.1 \lesssim g \lesssim 1$. It also explains why, in Eq. \eqref{sx}, we did not write the $\frac{1}{2}m_\chi \chi^2$ term, as the DM bare mass is heavily suppressed by the DM constraint and finally much smaller than the $\frac{1}{4}\lambda \chi^4$ term.

\subsection{The primordial black hole}
In this part, we want to describe the PBH production in Starobinsky inflation, especially in the preheating stage. This discussion has already been discussed in Ref. \cite{risdianto2026preheating}. However, the calculation in this paper is slightly different.

\begin{figure}
    \centering
    \includegraphics[width=0.7\linewidth]{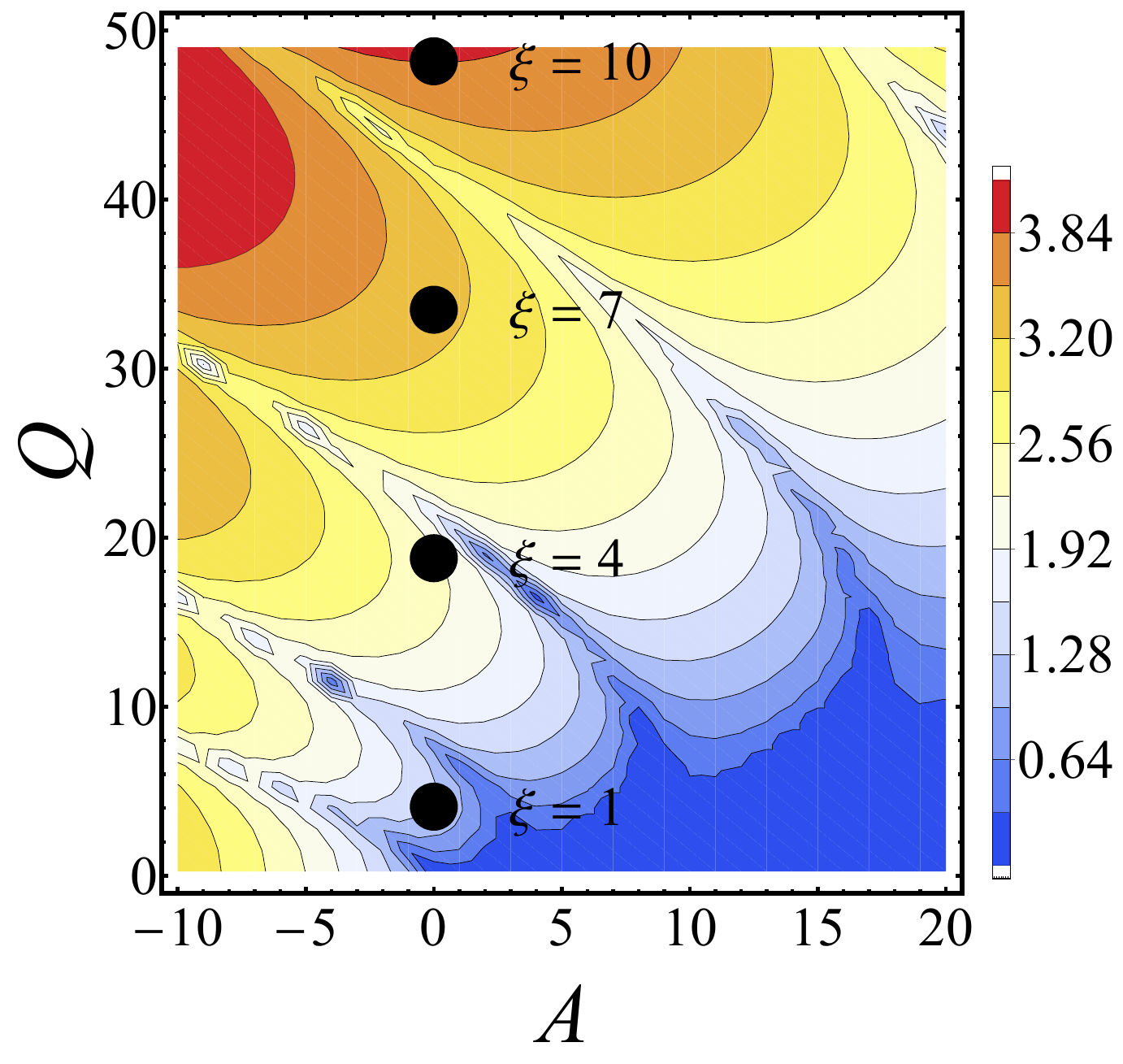}
    \caption{Mathieu instability chart; the color of the figure corresponds to the characteristic exponent $ \mu_K$ depicted on the right side of the figure.}
    \label{mathieu}
\end{figure}

Straightforwardly, we will write Eq. \eqref{ddotchik} as
the Mathieu equation as
\begin{equation}
     \frac{d^2 \chi_k}{dz^2}+[A-2Q\cos(2z)]\chi_k=0.
\end{equation}
where

\begin{equation}
    A= 4\left( \frac{k_\chi^2+3\lambda\braket{\chi^2}}{M^2}\right), \hspace{1cm} Q= \sqrt{\frac{2}{3}}\frac{6}{M^2}\left(\lambda\braket{\chi^2}+\left( \xi-\frac{1}{6}\right)M^2 \right)\frac{\tilde{\phi}}{M_p}
\end{equation}
The growth of $\chi_k$ is depicted by the characteristic exponent as
\begin{equation}
    \mu_K=\frac{1}{2\pi}\ln \left(1+2e^{-\pi K^2}-2\sin\vartheta e^{-\pi K^2} \sqrt{1-2e^{-\pi K^2}} \right)
\end{equation}
It is straightforward to write the $ \xi$-dependent Mathieu characteristic exponent in Fig. \ref{mathieu}. 

\begin{figure}
    \centering
    \includegraphics[width=0.7\linewidth]{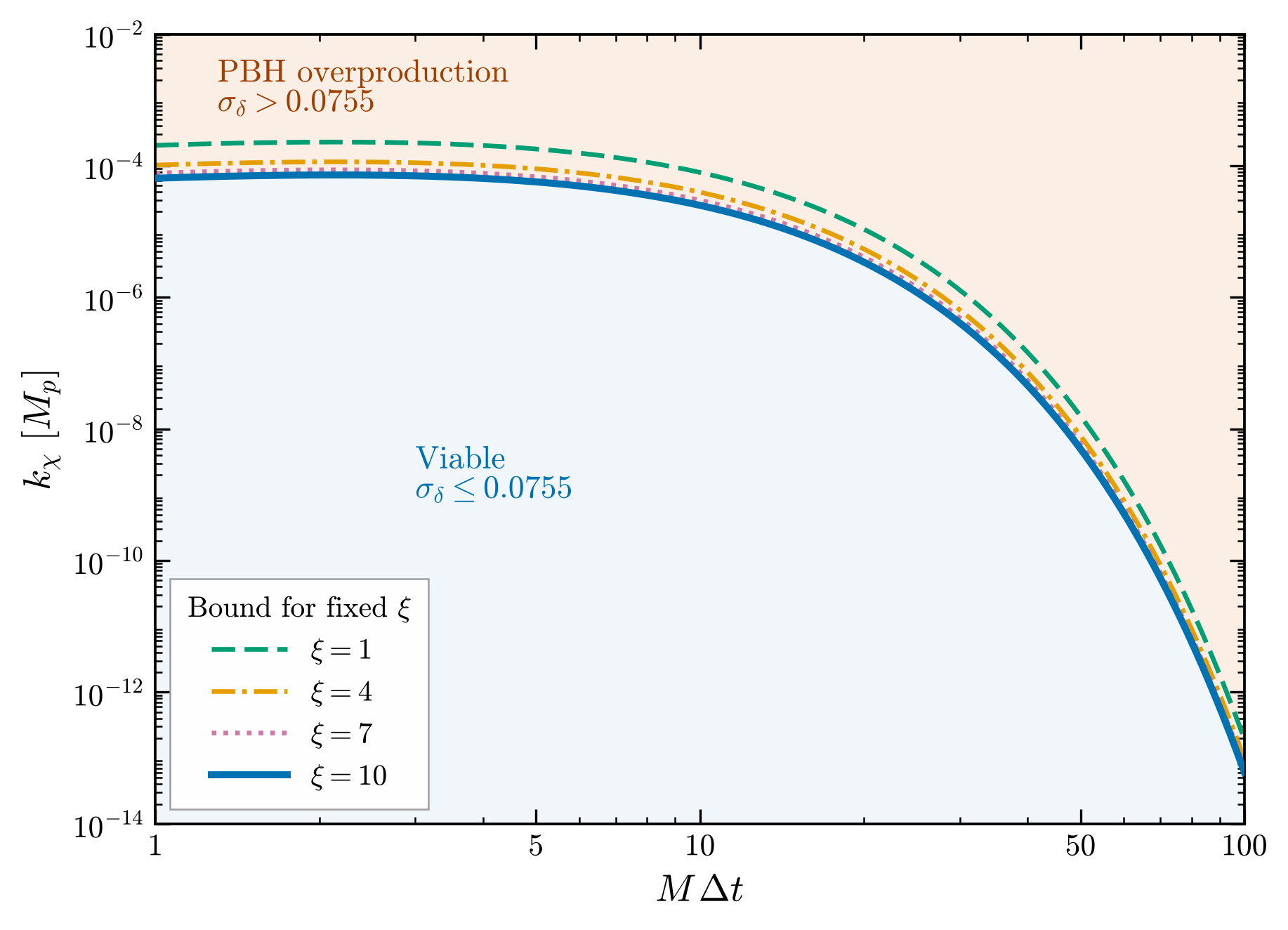}
    \caption{Plot of momentum $k_\chi$ versus the PBH duration $M\Delta t$. The variation of $\xi$ gives the weak constraint on the PBH duration. }
    \label{kchideltat}
\end{figure}

The constraint on the PBH can be written as
\begin{equation}\label{beta}
    \beta=\int^\infty_{\delta_c}P(\delta)d\delta,
\end{equation}
where $\beta=10^{20}$ and $\delta_c=0.7$ \cite{green2001primordial,niemeyer1999dynamics}. $P(\delta)$ corresponds to the probability distribution function as
\begin{equation}\label{pdelta}
    P(\delta)=\frac{1}{\sqrt{2\pi} \sigma_\delta} e^{ -\frac{\delta^2}{2\sigma_\delta^2}}.
\end{equation}
By evaluating Eqs. \eqref{beta} and \eqref{pdelta}, we obtain the mass variance $\sigma_\delta\simeq 0.08$. We will use this result later to constrain PBH production. The power spectrum of $\chi$ fluctuations generated during preheating can be estimated as
\begin{equation}
\mathcal{P}_{\delta\chi}|_\text{pre}=\mathcal{P}_{\delta\chi}|_\text{end}\exp(2\mu_K M\Delta t),
\end{equation}
where we take $\mu_K\approx \mu_0\simeq \frac{1}{2\pi}\ln 3$ as the characteristic exponent evaluated at the largest possible value of $\mu_K$, and $\Delta t$ is the time required to produce the desired PBH abundance.
On the other hand, we write

\begin{equation}
K^2=\frac{k_\chi^2+3\lambda\braket{\chi^2}}{ \left[3\sqrt{\frac{2}{3}}\frac{1}{M_p}\left(\lambda\braket{\chi^2}
    +\left(\xi-\frac{1}{6}\right) M^2  \right)\tilde{\phi} M\right]^{-2/3}} \equiv \frac{1}{18\sqrt{Q}}\left(\frac{k_\chi^2 +3\lambda\braket{\chi^2}}{k_\text{end}^2}\right)
\end{equation}
where

\begin{equation}
k_\text{end}^2=\frac{1}{18\sqrt{2}}\left(3\left[\sqrt{\frac{2}{3}}\left(\lambda\braket{\chi^2}+\left(\xi-\frac{1}{6}\right)M^2  \right) \frac{\Tilde{\phi}}{M_p}\right]^{1/6}\right)M^{5/3},
\end{equation}
is the comoving wavenumber evaluated at the Hubble exit at the end of inflation. The non-adiabatic power spectrum of $\chi$ can then be described as

\begin{equation}
    \mathcal{P}_{\zeta_\text{n-ad}}(k_\chi)\simeq \mathcal{A} \left(\frac{k_\chi}{k_\text{end}} \right)^3 I(K, M\Delta t),
\end{equation}
where\footnote{The third bracket is coming from the interaction coupling between $\phi$ and $\chi$ via  $\mathcal{L}_\text{int}=3\sqrt{\frac{3}{2}}\frac{\xi M^2}{M_p}\phi \chi^2$ divided by mass scale $M$.}

\begin{equation}
        \mathcal{A}\equiv\frac{3 \times 2^{3/2}}{\pi^6 \mu_p^2}\left(\frac{\Tilde{\phi}_\text{end}}{M_p} \right)^2\left(\frac{H_\text{end}}{M}\right)^4 \left(3\sqrt{\frac{3}{2}}\frac{\xi M}{M_p}\right)^2 Q^{-1/4}.
\end{equation}
Otherwise we have

\begin{equation}
\begin{split}
        I(K, M \Delta t)&\equiv \frac{3}{2}\int^{\infty}_0 K'^2dK'\int^\pi_0 \sin\theta d\theta e^{2(\mu_{K'}-\mu_{K-K'})M\Delta t}, \\
        &=\frac{3}{2}\int^{\infty}_0 K'^2dK'\int^\pi_0 \sin\theta d\theta  e^{4\mu_{K'}M\Delta t},
\end{split}
\end{equation}
where $\theta$ is the angle between $K$ and $K'$.
To solve the last equation, we need to impose \cite{green2001primordial,liddle2000super}
\begin{equation}
    \mu_{K-K'}\simeq \mu_K'+\frac{2K'\cos \theta}{2+e^{\pi K'^2}}+\mathcal{O}(K^2), \hspace{1cm}\mu_{K'} \approx \mu_0+\frac{K'^2}{3}.
\end{equation}
Straightforwardly, we obtain
\begin{equation}
\begin{split}
       I(K, M\Delta t)&\simeq\frac{3}{2}\int^{\infty}_0 K'^2dK'\int^\pi_0 \sin\theta d\theta  e^{4\mu_{K'}M\Delta t}\\
       &\simeq 0.863(M\Delta t)^{-3/2}\cdot e^{4\mu_{0}M\Delta t}.\\  
\end{split}
\end{equation}
Finally, the mass variance $\sigma_\delta$ during the matter-dominated era can be calculated as follows \cite{green2004NewCalculation}
\begin{equation}\label{sigmadelta}
    \sigma_\delta^2 =\frac{1}{2}\int^\infty_0 W^2(\Tilde{k}_\chi,\mathcal{R})\left( \frac{\Tilde{k}_\chi}{{k}_\chi}\right)^4 \mathcal{P}_{\zeta_\text{n-ad}}\frac{d\Tilde{k}_\chi}{\Tilde{k}_\chi},
\end{equation}
where we introduced the window function $W( \Tilde{k}_\chi,\mathcal{R})=\exp \left(  -\Tilde{k}_\chi^2 \mathcal{R}^2/2 \right)$, with $\mathcal{R}=1/k_\chi$.
Finally,  we can use  $\sigma_\delta\simeq 0.00755$ and calculation from Eqs. \eqref{sigmadelta} and \eqref{beta} to obtain the constraint on PBH production to be
\begin{equation}
    0.0755^2\gtrsim \frac{3026.97 \cdot e^{0.6998 \cdot M\Delta t}}{(M\Delta t)^{3/2}}\cdot   \frac{\xi^{3/2}k_\chi^3}{M M_p^2}.
\end{equation}
Figure \ref{kchideltat} illustrates the relationship between the momentum mode $k_\chi$ and the PBH production duration $\Delta t$. As shown in the figure, the duration of PBH production can be estimated for $M\Delta t\lesssim 100$.

\subsection{The end of the \textit{first} preheating and PBH duration}\label{endfirstpre}

For small values of $\Tilde{\phi}$ ($\ll M_p$), the term containing $\phi/M_p$ is suppressed, and Eq. \eqref{ddotchi} reduces to a simple harmonic oscillator, in which the backreaction effect eventually terminates the (first) preheating stage. In the following, we estimate the field amplitude $\chi_k$ at which preheating terminates due to backreaction.\footnote{In general, $\braket{\chi^2_k} \neq \chi_k^2$ because the variance involves an integral over all momentum modes. However, in the narrow resonance regime, particle production occurs only within a very narrow band of momenta. Consequently, a single mode (or a narrow band around $k_*$) dominates the dynamics, allowing us to approximate $\braket{\chi^2_k} \approx \chi_k^2$.}

\begin{equation}\label{narrowconditions}
    \braket{{\chi}_k^2}\approx \Tilde{\chi}_k^2 \gtrsim \frac{\xi M^2}{\lambda} \frac{\Tilde{\phi}_\text{crit}}{M_p}.
\end{equation}
$\tilde{\phi}_\text{crit}$ corresponds to the scalaron field value at which the spectator field takes over the preheating process. Before we proceed, we first obtain a rough estimate of $\tilde{\phi}_\text{crit}$. In the Einstein frame, Eq. \eqref{sechi} clearly shows that the potential of $\chi$ is suppressed by the factor $e^{-2\sqrt{\frac{2}{3}}\frac{\phi}{M_p}}$. When the spectator field takes over the preheating process, this suppression is lifted as $\tilde{\phi}$ becomes small. Consequently, $\Omega\sim 1$, and the actions in the Jordan and Einstein frames become approximately equivalent. This condition has been discussed in Refs. \cite{risdianto2025inflation,risdianto2025second}. Using the result of Ref. \cite{risdianto2025second}, we point out that the first preheating stage is expected to end when $\tilde{\phi}=\tilde{\phi}_\text{crit}\sim M$. Putting $\tilde{\phi}_\text{crit}=M$, we obtain
\begin{equation}\label{chibackreaction}
    \Tilde{\chi}_k\sim \sqrt{\frac{\xi M^3}{\lambda M_p}},
\end{equation} 
represents the initial conditions of the second preheating.

Regarding the duration of preheating, the calculation presented in this paper has also been used in Refs. \cite{hashimoto2021inflation,risdianto2025inflation}. The approximate duration of preheating can be explicitly derived from Eq. \eqref{phi} as
\begin{equation}\label{duration}
    \Delta t= t_\text{crit}-t_\text{end} =2\sqrt{\frac{2}{3}}\frac{M_p}{M}=2\sqrt{\frac{2}{3}}\frac{M_p}{M}\left(\frac{1}{\Tilde{\phi}_\text{crit}}-\frac{1}{\Tilde{\phi}_\text{end}}\right)\simeq 2\sqrt{\frac{2}{3}}\frac{M_p}{M \Tilde{\phi}_\text{crit}},
\end{equation}
where $t_\text{crit}$ and $\phi_\text{crit}$ denote the time and the value of the field $\phi$ at the critical point, marking the end of the first preheating stage. Similarly, $t_\text{end}$ and $\phi_\text{end}$ denote the time and the value of $\phi$ at the end of inflation. Taking $\tilde{\phi}_\text{crit}=M$ in Eq. \eqref{duration}, we find that the duration of the first preheating stage is approximately 3 orders longer than the PBH production duration. We also note that Ref. \cite{green2001primordial} suggests that PBH formation should occur during the early stages of preheating. As the universe expands rapidly, the subsequent expansion can suppress the collapse of perturbations into PBHs.

\section{The \textit{second} preheating}\label{secondpre}
\subsection{The particle production}
After backreaction takes over and preheating driven by the $\phi$ oscillations ceases, the $\chi_k$ field becomes the dominant driver of preheating. At this stage, the Ricci scalar in the Jordan frame satisfies $R_J\simeq R\sim M^3/M_p$.

In this stage, the equation of motion of $\chi$ can be approximated from the $S_\chi$ Eq. \eqref{se}  as follows
\begin{equation}
\Ddot{\chi}+3H\Dot{\chi}+3\lambda \chi^3=0.
\end{equation}
In addition, we used the conformal field as
\begin{equation}
    X=a\chi=a\tilde{\chi}f=\tilde{X}f,
\end{equation}
and the conformal time $d\tau=\sqrt{\lambda}\tilde{\chi}\frac{dt}{a}$. Finally, we obtain
\begin{equation}
    \frac{d^2f}{d\tau^2}+f^3=0.
\end{equation}
The solution is a Jacobi elliptic function, which can simply be written as \cite{kofman1997towards}
\begin{equation}
    f=\text{cn} \left( \tau, 1/\sqrt{2}\right).
\end{equation}
Before proceeding further, we introduce the Lagrangian density for the Higgs field:
\begin{equation}
    -\frac{1}{2}\partial_\mu h\partial_\nu h-\frac{1}{4}g\chi^2h^2-\frac{1}{4}\lambda_h h^4,
\end{equation}
where $g$ denotes the coupling between the Higgs field and $\chi$. At the beginning of the second preheating stage, the quartic Higgs potential, $\frac{1}{4}\lambda_h h^4$, is negligible. The equation of motion for the Higgs field can then be written as
\begin{equation}
    \Ddot{h}_k+3H\Dot{h}_k+\left(\frac{k_h^2}{a^2}+\frac{1}{2}g\chi^2 \right)h_k=0,
\end{equation}
where $h_k$ is defined using the Heisenberg representation, as in Eq. \eqref{heisenberg}. We then defining
\begin{equation}
    \mathcal{H}\equiv ah_k, \hspace{1cm}\kappa_h\equiv \frac{k_h}{a\sqrt{\lambda}\tilde{\chi}},.
\end{equation}
Finally we obtain
\begin{equation}
    \frac{d^2\mathcal{H}}{d\tau^2}+\left(\kappa_h+\frac{g}{2\lambda}\text{cn}^2\left(\tau, \frac{1}{\sqrt{2}}\right)\right)\mathcal{H}=0.
\end{equation}
The conformal energy density produced by the single crossing can be estimated at \cite{risdianto2025second}
\begin{equation}\label{quarticenergy}
\delta \Tilde{\rho}_h\approx\int^\infty_0\frac{d^3{k}_h}{(2\pi)^3}\exp\left(-\pi\frac{k_h^2}{g/4\lambda}\right)\sqrt{\frac{g}{2\lambda}}\simeq 1.786 \times 10^{-4} \hspace{1mm}\frac{g^2}{\lambda^2},
\end{equation}
which corresponds to the physical energy density (neglecting the expansion of the universe) as 
\begin{equation}
    \delta \rho_h =  \delta \Tilde{\rho}_h\times \lambda^2\Tilde{\chi}^4  =1.786 \times 10^{-4} \hspace{1mm}\frac{g^2}{\lambda^2} \times \rho_\chi,
\end{equation}
where $\rho_\chi=\frac{1}{4}\lambda \Tilde{\chi}^4$ is the remaining spectator's field energy density. As $1.786 \times 10^{-4} \hspace{1mm}\frac{g^2}{\lambda^2}\gg 1$, it is clear that during the second preheating is ceased after only one crossing, making this mode is extremely short.

\subsection{The reheating temperature}\label{reheating}
In this subsection, we will discuss further the reheating temperature in Starobinsky inflation, which, to be precise, is mostly determined by the spectator field $\chi$. It has been discussed in Ref. \cite{risdianto2025second} that the reheating temperature by the simple decay is disfavored. In our model, when the second preheating approaches, the $\chi$ field becomes the sole contributor to the preheating, replacing the scalaron. The $\chi$ field is oscillating and non-perturbatively decays to the Higgs boson. Finally, its decay dominant channel to $\Bar{b}b$ will reheat the universe. This mechanism is already discussed in Ref. \cite{risdianto2025second,risdianto2025inflation, risdianto2026preheating}. Straightforwardly, we need to calculate the conformal energy density of the produced Higgs from the $\chi$ first-crossing followed by the instantaneous decay to $\Bar{b}b$ as \cite{ballesteros2017standard}
\begin{equation}\label{deltabarrhoh}
    \delta \Bar{\rho}_h=\int^{\tau_0}_0 \Bar{\Gamma}_{h\rightarrow\Bar{b}b}(\tau)\Bar{m}_h\Bar{n}_h e^{-\int^\tau_0 \Bar{\Gamma}_h (\tau')d\tau'}d\tau,
\end{equation}
where $\tau_0=7.416$ is the
period of the oscillations \cite{greene1997structure}. In solving Eq. \eqref{deltabarrhoh}, we need to write the conformal mass as
\begin{equation}
    \Bar{m}_h=\sqrt{\frac{g}{2\lambda}}\sin(c\tau),
\end{equation}
where we used $c=2\pi/\tau_0$, the conformal number density as
\begin{equation}
    \Bar{n}_h=\int^\infty_0 \frac{d^3k_h}{(2\pi)^3}e^{-\pi \frac{k_h^2}{g/2\lambda}}= \frac{1}{8\pi^3}\left( \frac{g}{2\lambda}\right)^{3/2}.
\end{equation}
The conformal decay of the Higgs to $\Bar{b}b$ can be written as
\begin{equation}
    \bar{\Gamma}_{h\rightarrow \bar{b}b}\approx\frac{3y_b^2\Bar{m}_h}{16\pi}= \frac{3y_b^2 }{16\pi}\sqrt{\frac{g}{2\lambda}}\sin(c\tau),
\end{equation}
where $y_b$ is the Yukawa coupling between Higgs and bottom quark. Straightforwardly, we can obtain

\begin{equation}
    \delta \bar{\rho}\approx \frac{y_b^2\tau_0}{256\pi^4}\left(\frac{g}{2\lambda}\right)^{5/2},
\end{equation}
the energy density transferred to the bottom quarks is averaged and estimated as
\begin{equation}
    \bar{\rho}=\frac{2\delta \bar{\rho}}{\tau_0}\tau
\end{equation}
The conformal energy density takes the form of $\bar{\rho}=\frac{1}{4\lambda}$. We can convert it to physical energy density and straightforwardly connect it to the  reheating temperature via
\begin{equation}
    \frac{1}{4\lambda}\left( \frac{\sqrt{\lambda} \Tilde{\chi} }{a}\right)^4=\frac{g^* \pi}{30} T_R^4.
\end{equation}
Finally, by using $y_b=0.024$, $g^*=106.75$ as the standard model degrees of freedom, and $a=\frac{\tilde{\chi}}{2\sqrt{3}M_p}\tau$, we obtain
\begin{equation}
T_R \simeq 2.66\times10^{10} \sqrt{\frac{g^5}{\lambda^3 }}\text{GeV}.
\end{equation}
This gives a strong constraint on $\lambda$. In Eq. \eqref{mchig} it shows the constraint on $g$ is $0.1\lesssim g\lesssim 1$. If the reheating temperature's upper bound from the ACT results is $T_R\sim 10^{15}$ GeV, we obtain
\begin{equation}
    \lambda \lesssim 10^{-3}.
\end{equation}

\section{Conclusion}
In this paper, we investigate the post-inflationary predictions of the Starobinsky model in light of the new ACT results. We retain the conventional Starobinsky inflationary scenario, which assumes a post-inflationary equation of state $w=0$. Under this assumption, the model lies at the edge of the ACT-LB results but is excluded by the P-ACT-LB constraints. We consider this scenario because the conventional Starobinsky potential becomes approximately quadratic near its minimum, naturally leading to a matter-dominated post-inflationary regime. This provides a basis for investigating whether the conventional Starobinsky model remains compatible with the new observational results while maintaining a consistent post-inflationary and preheating scenario. By contrast, allowing $w\neq 0$ would require substantial modifications to the post-inflationary dynamics of the Starobinsky model to accommodate the observational constraints. For instance, $w\simeq 0.5$, which is favored by the P-ACT-LB results according to Ref. \cite{zharov2025act}, would substantially alter the preheating dynamics and potentially require an extension of the conventional Starobinsky model.

In this paper, we divide the preheating stage into two stages, following the approach adopted in Refs. \cite{risdianto2025second,risdianto2025inflation}. During the first preheating stage, inflaton oscillations give rise to several important phenomena, including tachyonic preheating, which drives efficient energy transfer, as well as PBH formation and DM production. Since the scalaron potential is approximately quadratic near its minimum, the universe undergoes a matter-dominated phase. Subsequently, when the scalaron field decreases to $M$, the spectator field takes over the post-inflationary dynamics, marking the beginning of the second preheating stage. During this stage, the quartic potential of $\chi$ drives the universe toward a radiation-dominated regime. Consequently, the second preheating stage plays a crucial role in determining the reheating temperature.

During the first preheating stage, efficient preheating in Starobinsky inflation depends strongly on the initial conditions of the spectator field $\chi$ and the non-minimal coupling $\xi$. Our analytical results indicate that energy transfer is primarily driven by tachyonic preheating and can drain the entire inflaton energy density only for $\xi \gtrsim 100$. A more detailed numerical analysis is required to determine the precise particle production dynamics, which is beyond the scope of the present work. We also consider $\chi$ as a DM candidate and derive a constraint on the coupling $g$, which is found to be of order unity. Furthermore, our analysis indicates that PBHs are predominantly produced during the very early stages of preheating,\footnote{This is consistent with our analysis, in which the universe is dominated by heavy matter immediately after inflation, naturally favoring $w=0$.} over a production duration much shorter than the total preheating duration. Thus, our results do not contradict the possibility of PBH formation during preheating \cite{green2001primordial}. We also find that PBH formation depends strongly on the comoving momentum mode but only weakly on the non-minimal coupling $\xi$.

Finally, the reheating temperature inferred from the ACT-LB results is expected to be of the order of $10^{14}\sim 10^{15}$ GeV, which is significantly higher than that inferred from the Planck results. The reheating mechanism is directly associated with the second preheating stage, which is driven by the oscillating spectator field. We estimate the reheating temperature from the first zero crossing of the spectator field after the onset of the second preheating stage. During this crossing, Higgs bosons are produced through their direct coupling to the spectator field. Subsequently, the Higgs bosons predominantly decay into $b\bar{b}$ pairs, thereby reheating the universe.

\section*{Acknowledgment}
M.A.S. would like to thank the Indonesia Endowment Fund for Education (LPDP) Scholarship Program, Ministry of Finance, Republic of Indonesia.

\bibliographystyle{JHEP}
\bibliography{name}

@article{louis2025atacama,
title={The Atacama Cosmology Telescope: DR6 Power Spectra, 
   Likelihoods and $\Lambda$ CDM Parameters},
  author={Louis, Thibaut and La Posta, Adrien and Atkins, Zachary and Jense, Hidde T and Abril-Cabezas, Irene and Addison, Graeme E and Ade, Peter AR and Aiola, Simone and Alford, Tommy and Alonso, David and others},
  journal={arXiv preprint arXiv:2503.14452},
  year={2025}
}

@article{risdianto2026preheating,
  title={The preheating stage on the Starobinsky inflation after ACT},
  author={Risdianto, Norma Sidik and Budhi, Romy Hanang Setya and Shobcha, Nehla and Salim Adam, Apriadi and Syakura, Muhammad Abdan},
  journal={Physica Scripta},
  volume={101},
  number={36},
  pages={365003},
  year={2026},
  publisher={IOP Publishing}
}

@article{de2010f,
  title={f (R) theories},
  author={De Felice, Antonio and Tsujikawa, Shinji},
  journal={Living Reviews in Relativity},
  volume={13},
  number={1},
  pages={1--161},
  year={2010},
  publisher={Springer}
}

@article{risdianto2025second,
  title={The second-level preheating},
  author={Risdianto, Norma Sidik and Adam, Apriadi Salim and Zamakhsyari, Lalu},
  journal={Journal of Cosmology and Astroparticle Physics},
  volume={2025},
  number={01},
  pages={109},
  year={2025},
  publisher={IOP Publishing}
}

@article{felder1999inflation,
  title={Inflation and preheating in NO models},
  author={Felder, Gary and Kofman, Lev and Linde, Andrei},
  journal={arXiv preprint hep-ph/9903350},
  year={1999}
}

@article{risdianto2025inflation,
  title={On Inflation and Reheating Features in the Higgs-R2 Model},
  author={Risdianto, Norma Sidik},
  journal={Nuclear Physics B},
  pages={117149},
  year={2025},
  publisher={Elsevier}
}

@article{ballesteros2017standard,
  title={Standard Model—axion—seesaw—Higgs portal inflation. Five problems of particle physics and cosmology solved in one stroke},
  author={Ballesteros, Guillermo and Redondo, Javier and Ringwald, Andreas and Tamarit, Carlos},
  journal={Journal of Cosmology and Astroparticle Physics},
  volume={2017},
  number={08},
  pages={001--001},
  year={2017}
}

@article{niemeyer1999dynamics,
  title={Dynamics of primordial black hole formation},
  author={Niemeyer, Jens C and Jedamzik, Karsten},
  journal={Physical Review D},
  volume={59},
  number={12},
  pages={124013},
  year={1999},
  publisher={APS}
}

@article{liddle2000super,
  title={Super-horizon perturbations and preheating},
  author={Liddle, Andrew R and Lyth, David H and Malik, Karim A and Wands, David},
  journal={Phys. Rev. D},
  volume={61},
  number={10},
  pages={103509},
  year={2000},
  publisher={APS}
}

@article{terada2015reheating,
  title={Reheating processes after Starobinsky inflation in old-minimal supergravity},
  author={Terada, Takahiro and Watanabe, Yuki and Yamada, Yusuke and Yokoyama, Jun’ichi},
  journal={Journal of High Energy Physics},
  volume={2015},
  number={2},
  pages={1--26},
  year={2015},
  publisher={Springer}
}

@article{dorsch2024gravitational,
  title={Gravitational reheating in Starobinsky inflation},
  author={Dorsch, Gl{\'a}uber C and Miranda, Luiz and Yokomizo, Nelson},
  journal={Journal of Cosmology and Astroparticle Physics},
  volume={2024},
  number={11},
  pages={050},
  year={2024},
  publisher={IOP Publishing}
}

@techreport{dolgov1989particle,
  title={On particle creation by a time-dependent scalar field},
  author={Dolgov, AD and Kirilova, DP and others},
  year={1989},
  institution={Joint Inst. for Nuclear Research, Dubna (USSR). Lab. of Theoretical Physics}
}

@article{dufaux2007theory,
  title={Theory and numerics of gravitational waves from preheating after inflation},
  author={Dufaux, Jean-Fran{\c{c}}ois and Bergman, Amanda and Felder, Gary and Kofman, Lev and Uzan, Jean-Philippe},
  journal={Physical Review D—Particles, Fields, Gravitation, and Cosmology},
  volume={76},
  number={12},
  pages={123517},
  year={2007},
  publisher={APS}
}

@article{dufaux2010gravitational,
  title={Gravitational waves from Abelian gauge fields and cosmic strings at preheating},
  author={Dufaux, Jean-Francois and Figueroa, Daniel G and Garcia-Bellido, Juan},
  journal={Physical Review D—Particles, Fields, Gravitation, and Cosmology},
  volume={82},
  number={8},
  pages={083518},
  year={2010},
  publisher={APS}
}

@article{bethke2013anisotropies,
  title={Anisotropies in the gravitational wave background from preheating},
  author={Bethke, Laura and Figueroa, Daniel G and Rajantie, Arttu},
  journal={Physical review letters},
  volume={111},
  number={1},
  pages={011301},
  year={2013},
  publisher={APS}
}

@article{garcia2007stochastic,
  title={Stochastic background of gravitational waves from hybrid preheating},
  author={Garcia-Bellido, Juan and Figueroa, Daniel G},
  journal={Physical review letters},
  volume={98},
  number={6},
  pages={061302},
  year={2007},
  publisher={APS}
}

@article{hashimoto2021inflation,
  title={Inflation connected to the origin of CP violation},
  author={Hashimoto, Tsuyoshi and Risdianto, Norma Sidik and Suematsu, Daijiro},
  journal={Physical Review D},
  volume={104},
  number={7},
  pages={075034},
  year={2021},
  publisher={APS}
}

@article{kofman1994reheating,
  title={Reheating after inflation},
  author={Kofman, Lev and Linde, Andrei and Starobinsky, Alexei A},
  journal={Physical Review Letters},
  volume={73},
  number={24},
  pages={3195},
  year={1994},
  publisher={APS}
}

@article{shtanov1995universe,
  title={Universe reheating after inflation},
  author={Shtanov, Yu and Traschen, J and Brandenberger, RH1995PhRvD},
  journal={Physical Review D},
  volume={51},
  number={10},
  pages={5438},
  year={1995},
  publisher={APS}
}

@article{traschen1990particle,
  title={Particle production during out-of-equilibrium phase transitions},
  author={Traschen, Jennie H and Brandenberger, Robert H},
  journal={Physical Review D},
  volume={42},
  number={8},
  pages={2491},
  year={1990},
  publisher={APS}
}

@article{mohammadi2026starobinsky,
    author = "Mohammadi, Abolhassan and Yogesh and Tan, Hongwei and Sami, M.",
    title = "{A Starobinsky-like Potential in k-Essence Framework: Attractor Dynamics, Reheating, and Consistency with ACT DR6}",
    eprint = "2607.06112",
    archivePrefix = "arXiv",
    primaryClass = "gr-qc",
    journal = "Phys. Dark Univ.",
    volume = "54",
    pages = "102489",
    year = "2026"
}

@article{sotiriou2010f,
  title={f (R) theories of gravity},
  author={Sotiriou, Thomas P and Faraoni, Valerio},
  journal={Reviews of Modern Physics},
  volume={82},
  number={1},
  pages={451--497},
  year={2010},
  publisher={APS}
}

@article{rodrigues2022higher,
  title={Higher-order extension of Starobinsky inflation: Initial conditions, slow-roll regime, and reheating phase},
  author={Rodrigues-da-Silva, G and Bezerra-Sobrinho, J and Medeiros, LG},
  journal={Physical Review D},
  volume={105},
  number={6},
  pages={063504},
  year={2022},
  publisher={APS}
}

@article{drees2025refined,
  title={Refined predictions for Starobinsky inflation and post-inflationary constraints in light of ACT},
  author={Drees, Manuel and Xu, Yong},
  journal={Physics Letters B},
  volume={867},
  pages={139612},
  year={2025},
  publisher={Elsevier}
}

@article{zharov2025act,
  title={ACT observations, reheating, and Starobinsky and Higgs inflation},
  author={Zharov, DS and Sobol, OO and Vilchinskii, SI},
  journal={Physical Review D},
  volume={112},
  number={2},
  pages={023544},
  year={2025},
  publisher={APS}
}

@article{van2017reheating,
  title={Reheating and preheating in the simplest extension of Starobinsky inflation},
  author={van de Bruck, Carsten and Dunsby, Peter and Paduraru, Laura E},
  journal={International Journal of Modern Physics D},
  volume={26},
  number={13},
  pages={1750152},
  year={2017},
  publisher={World Scientific}
}

@article{starobinsky1980new,
  title={A new type of isotropic cosmological models without singularity},
  author={Starobinsky, Alexei A},
  journal={Physics Letters B},
  volume={91},
  number={1},
  pages={99--102},
  year={1980},
  publisher={Elsevier}
}

@article{garcia2023reheating,
  title={Reheating constraints and consistency relations of the Starobinsky model and some of its generalizations},
  author={Garcia, Marcos AG and Germ{\'a}n, Gabriel and Gonzalez Quaglia, R and Moran Colorado, AM},
  journal={Journal of Cosmology and Astroparticle Physics},
  volume={2023},
  number={12},
  pages={015},
  year={2023},
  publisher={IOP Publishing}
}

@article{akrami2020planck,
  title={Planck 2018 results-X. Constraints on inflation},
  author={Akrami, Yashar and Arroja, Frederico and Ashdown, M and Aumont, J and Baccigalupi, Carlo and Ballardini, M and Banday, Anthony J and Barreiro, RB and Bartolo, Nicola and Basak, S and others},
  journal={Astronomy \& Astrophysics},
  volume={641},
  pages={A10},
  year={2020},
  publisher={EDP sciences}
}

@article{kehagias2014remarks,
  title={Remarks on the Starobinsky model of inflation and its descendants},
  author={Kehagias, Alexandros and Moradinezhad Dizgah, Azadeh and Riotto, Antonio},
  journal={Physical Review D},
  volume={89},
  number={4},
  pages={043527},
  year={2014},
  publisher={APS}
}

@article{green2004NewCalculation,
  title={New calculation of the mass fraction of primordial black holes},
  author={Green, Anne M and Liddle, Andrew R and Malik, Karim A and Sasaki, Misao},
  journal={Phys. Rev. D},
  volume={70},
  number={4},
  pages={041502},
  year={2004},
  publisher={APS}
}

@article{green2001primordial,
  title={Primordial black hole production due to preheating},
  author={Green, Anne M and Malik, Karim A},
  journal={Phys. Rev. D.},
  volume={64},
  number={2},
  pages={021301},
  year={2001},
  publisher={APS}
}

@article{greene1997structure,
  title={Structure of resonance in preheating after inflation},
  author={Greene, Patrick B and Kofman, Lev and Linde, Andrei and Starobinsky, Alexei A},
  journal={Physical Review D},
  volume={56},
  number={10},
  pages={6175},
  year={1997},
  publisher={APS}
}

@article{kofman1997towards,
  title={Towards the theory of reheating after inflation},
  author={Kofman, Lev and Linde, Andrei and Starobinsky, Alexei A},
  journal={Physical Review D},
  volume={56},
  number={6},
  pages={3258},
  year={1997},
  publisher={APS}
}

\end{document}